%% file: main.tex
\documentclass[%
 reprint,
 superscriptaddress,
 amsmath,amssymb,
 aps,
 pra,
]{revtex4-1}

\usepackage{graphicx}
\usepackage{dcolumn}
\usepackage{bm}
\usepackage{hyperref}
\usepackage{url}
\usepackage{comment}

\newcommand{\eref}[1]{Eq.~(\ref{#1})}
\newcommand{\Eref}[1]{Equation~(\ref{#1})}
\newcommand{\fref}[1]{Fig.~\ref{#1}}
\newcommand{\Fref}[1]{Figure~\ref{#1}}

\newcommand{\Frefs}[1]{Figures~\ref{#1}}

\begin{document}
\newcommand{\ManuscriptTitle}{
    Conservation of Fractional Mean Energy
    in Dissipative Gases
}

\title{\ManuscriptTitle}


\author{Keisuke Fujii}
\email{fujiik@ornl.gov}
\affiliation{%
Oak Ridge National Laboratory, Oak Ridge, TN 37831-6305, United States of America
}

\date{\today}

\begin{abstract}
I show a nontrivial functional giving a conservation quantity in the collisional energy cascade of dissipative Maxwell gases: a fractional-calculus extension of the mean energy.
The conservation of this quantity directly leads the power-law energy tail that is stationary during the temporal evolution. 
In the thermal limit, this quantity naturally reduces to the standard mean energy.
%
This conservation law and its extension to particles with other interactions are demonstrated with a Monte-Carlo simulation for inelastic gases.
\end{abstract}

\maketitle

Among the variety of nonthermal dissipative systems, gaseous ensemble of inelastic particles has been extensively studied~\cite{Brilliantov2004-dj,Aranson2006-qf}.
Even with small inelasticity, inelastic gases show distinctive phenomena that are not seen in thermal systems.
For example, Ben-Naim et al. have shown a nontrivial steady-state distribution having a power-law energy tail under an extreme heating condition, where a particle with extremely high kinetic energy is injected at a certain rate into a isotropically distributed inelastic particles and this injected energy is balanced with the dissipation by the inelastic collisions~\cite{Ben-Naim2005-rd,Ben-Naim2005-uz,Kang2010-tk}. 
Such a power-law tail is a typical signature of nonthermal systems and
similar tails have been reported both in the natural~\cite{Bell1978-wg,Gutenberg1944-wb,Kolmogorov1991-uo,Hasegawa1977-oz} and social phenomena~\cite{Zipf1950-zy}.

In general, the transition from no-dissipation limit of nonthermal systems to thermal systems is not straightforward.
Indeed, Ben-Naim et al.~\cite{Ben-Naim2005-rd,Ben-Naim2005-uz,Kang2010-tk} predicts a finite power-law index for the energy tail even at the no-dissipation limit (i.e., elastic particles without heating), although the steady-state distribution in the thermal system is obviously the Maxwell distribution, the tail of which decays exponentially.
It has been also known that the energy spectrum of the fluid turbulence in thermal systems are completely different from that in dissipative systems even with infinitesimally small dissipation~\cite{Kraichnan1967-qf}.

In this Letter, for the heated inelastic gaseous system studied in Refs.~\cite{Ben-Naim2005-rd,Ben-Naim2005-uz,Kang2010-tk}, I show that an extension of mean energy based on fractional calculus is conserved, where instead of a normal integration to calculate the expectation (for the thermal system) the Riemann fractional integration is used.
This conservation law is equivalent with the time-invariant power-law tail in the energy distribution during the temporal evolution.
Furthermore, this definition reduces to the standard mean energy in the thermal system, which is obviously conserved.
This conservation law is demonstrated with numerical simulations
\footnote[1]{%
    The direct molecular dynamics simulation for more realistic systems, the details of the probabilistic representation of \eref{eq:recursive}, and the detailed derivation several equations can be found in Supplemental Material, which includes Refs.~\cite{Ito1985-px,Corrigan1965,Hey2004,McConkey2008}
}.

Establishment of thermodynamics for nonthermal systems has been a long-standing open question.
The conservation of the fractional mean energy may be interpreted as the first law of thermodynamics for this nonthermal system.


Let us consider an isotropic and spatially uniform ensemble of particles undergoing elastic collisions (i.e., no energy dissipation at this point) in $D$-dimensional space. 
We assume the Maxwell-type inter-particle interaction for now. Particle ensembles with other interactions will be discussed later. 
With the Maxwell interaction, the kinetic energies of two colliding particles, $E_1$ and $E_2$, can be thought as random samples from the energy distribution $f(E)$.
The following relation has been proposed for the post-collision energy $E_1'$~\cite{Futcher1980-ey,Hendriks1982-cw,Futcher1983-sf,Futcher1980-ey,Note1},
\begin{align}
  E_1' \leftarrow x E_1 + y E_2,
  \label{eq:recursive}
\end{align}
where $x, y \in[0,1]$ are random numbers following the probability distribution $p(x,y)$, which are determined by the collision geometry, such as the scattering angle and the relation between the relative and center-of-mass velocities. 
Several forms of $p(x, y)$ have been proposed. The simplest example of valid $p(x,y)$ is so-called \textit{diffuse} collision~\cite{Futcher1980-ey,Hendriks1982-cw}, where after the elastic collision the kinetic energies of the two particles will be completely randomized with the total energy conserved, i.e., no memory effect of pre-collision energies,
\begin{align}
p(x,y) = B\left( x \middle| \frac{D}{2}, \frac{D}{2}\right) \delta(x-y),
\label{eq:diffuse}
\end{align}
where $B(x|a, b) = x^{a-1} (1-x)^{b-1} / B(a, b)$ is beta distribution with beta function $B(a, b) = \int_0^1 x^{a-1} (1-x)^{b-1} dx$ and $\delta(t)$ is Dirac's delta function.
The $p$-$q$ model~\cite{Futcher1983-sf,Futcher1980-ey}, which takes the memory effect into account, as well as its linear superposition also give a valid $p(x,y)$~\cite{Note1}.

%
The temporal evolution \eref{eq:recursive} can be written in the following form with the Laplace transform of the energy distribution $\mathcal{L}_f(\tau, s) \equiv \int_0^\infty f(\tau, E)e^{-sE}dE$, 
\begin{align}
    \notag
    \frac{\partial}{\partial\tau}\mathcal{L}_f(\tau, s) 
    &= -\mathcal{L}_f(\tau, s) \\
    & + \int\, \mathcal{L}_f(\tau, xs) \mathcal{L}_f(\tau, ys)\, p(x, y) \,dx\,dy,
    \label{eq:laplace_nodissipation}
\end{align}
where $\tau$ is the time scaled by the collision frequency.
With any valid $p(x,y)$, the mean energy $\langle E\rangle \equiv\mathcal{L}_f'(\tau, 0) = \int E f(\tau, E) dE$ is conserved during the temporal evolution and eventually at the steady state the distribution converges to the Maxwell distribution $\mathcal{L}_f(\infty,s) = [1 + 2D^{-1}\langle E \rangle s]^{-D/2}$ according to Boltzmann's H-theorem.

Let us additionally consider an energy-dissipation.
We assume that, by this dissipation process, a particle looses its kinetic energy by the fraction of $1-e^{-\Delta}$ (with $\Delta \geq 0$).
We can assume an inelastic collision as this dissipation process, but other processes may be also considered.
The time evolution with this dissipation is
\begin{align}
    E_1' \leftarrow
    \begin{cases}
        e^{-\Delta} E_1, & \text{with probability}\ \xi\\
        x E_1 + y E_2, & \text{with probability}\ 1 - \xi
    \end{cases},
    \label{eq:recursive_dissipation}
\end{align}
where $\xi$ is the rate of this dissipation process relative to the elastic collision.
The Laplace representation of \eref{eq:recursive_dissipation} is
\begin{align}
    \notag
    \frac{\partial}{\partial\tau}&\mathcal{L}_f(\tau, s) =
    -\mathcal{L}_f(\tau, s) + \xi \mathcal{L}_f(\tau, e^{-\Delta}s)\\
    &+(1 - \xi) \int\, \mathcal{L}_f(\tau, xs) \mathcal{L}_f(\tau, ys)\, p(x, y) \,dx\,dy.
    \label{eq:laplace_dissipation}
\end{align}
Here, we implicitly assume a constant energy injection in the high energy limit so that the system will eventually arrive at a nontrivial steady state~\cite{Ben-Naim2005-rd}.

Let us consider the first two orders of $\mathcal{L}_f(\tau, s)$. 
From the normalization condition $\mathcal{L}_f(\tau, 0) = 1$, we may write $\mathcal{L}_f(\tau, s)\approx 1 - c(\tau) s^{\alpha(\tau)}$ in the small-$|s|$ region, with $0 < \alpha(\tau) \leq 1$.
Note that this corresponds to an assumption of $f(E)$ in the large-$E$ region, i.e., either $f(\tau, E) \approx E^{-\alpha-1} / c\Gamma(1-\alpha)$ if $\alpha < 1$, or $f(\tau, E) \approx \exp(-E/c)/c$ if $\alpha=1$.
In principle, $\alpha$ and $c$ can evolve in time. 
However, by substituting it to \eref{eq:laplace_dissipation}, comparing the terms with the orders of $s^{\alpha-1}$, $s^0$, and $s^{\alpha}$, we find
$d\alpha(\tau)/d\tau = 0.$
From the assumption that the system has a nontrivial steady state~\cite{Note1}, we then obtain $dc(\tau)/d\tau = 0$ and
\begin{align}
            1 &= (1-\xi) \int (x^\alpha + y^\alpha) p(x, y)\, dx\, dy + \xi e^{-\alpha \Delta}.
    \label{eq:alpha}
\end{align}
Note that the symmetry of the elastic collision leads $p(x,y) = p(1-y, 1-x)$, which results in $\int (x+y) p(x, y) dx\, dy = 1$~\cite{Note1}. 
This indicates that $\alpha = 1$ is the necessary and sufficient condition for the non-dissipative system, i.e., $\Delta = 0$ or $\xi = 0$.
With a finite energy dissipation, $\alpha$ is smaller than 1.

The above relations can be summarized as
\begin{align}
    \frac{\partial}{\partial\tau}\lim_{|s|\rightarrow 0} s^{1-\alpha} \frac{d \mathcal{L}_f(\tau,s)}{ds} = 0.
    \label{eq:small_s}
\end{align}
Let us define the generalized mean energy of this system as follows~\cite{Note1},
\begin{align}
    \notag
    \langle E \rangle_\alpha 
    &\equiv \left[
        \lim_{|s|\rightarrow 0} s^{1-\alpha} \frac{d \mathcal{L}_f(\tau, s)}{ds}\right]^{1/\alpha} \\
    &= \left[
        \lim_{E\rightarrow \infty} \frac{1}{\Gamma(\alpha)}
        \int_0^E (E-x)^{\alpha-1} x f(\tau, x) dx
    \right]^{1/\alpha},
    \label{eq:fractional_mean}
\end{align}
where $\Gamma(x)=\int_0^\infty t^{x-1}e^{-t}dt$ is the gamma function, and
the right hand side of \eref{eq:fractional_mean} involves the Riemann fractional integration of order $\alpha$~\cite{Herrmann2014-ve,Anatolii_Aleksandrovich_Kilbas2006-ts}. 
From \eref{eq:small_s}, we find that $\langle E \rangle_\alpha$ is conserved in this system, i.e., $d \langle E \rangle_\alpha / d \tau = 0$.
At the no-dissipation limit ($\alpha = 1$), $\langle E \rangle_\alpha$ reduces to the standard expectation $\langle E \rangle$ [note that the fractional integration of order 1 is the standard integration], which is a conserved quantity in the thermal system.
Thus, this is a conservation law for both the nonthermal and thermal systems.

The conservation of $\langle E \rangle_\alpha$ is equivalent with the time-invariant power-law tail 
$f(\tau, E)\approx (\langle E\rangle_\alpha)^{\alpha} E^{-\alpha-1}/\Gamma(1-\alpha)$ in the large $E$ region.
As this is time-invariant, the distribution has the same power-law tail at the steady state.
This is consistent with the argument by Ben-Naim et al.~\cite{Ben-Naim2005-rd,Ben-Naim2005-uz,Kang2010-tk}, where the steady-state velocity distribution of inelastic gases has a power-law tail, and the index of the power-law tail converges to a finite value (which is 2 for Maxwell gases) at the no-dissipation limit ($\alpha\rightarrow 1$). 
Furthermore, our theory explains how this tail converges to the Maxwell distribution at the thermal limit [observe that $\Gamma(\epsilon)^{-1} \approx \epsilon$ with $0 < \epsilon \ll 1$].

At the large-$s$ limit, $\mathcal{L}_f(\infty, s)$ asymptotically behaves $\approx 2D^{-1}(\langle E \rangle_\alpha s)^{-D/2}$ at the steady state if $\Delta \ll 1$~\cite{Note1}.
By combining with the lowest order approximation 
$\mathcal{L}_f(\tau, s) \approx 1 - (\langle E \rangle_\alpha s)^\alpha$, 
we find that the generalized Mittag-Leffler (GML) distribution~\cite{haubold_mittag-leffler_2011,barabesi_new_2016,korolev_mixture_2020},
\begin{align} 
    \mathcal{L}_f(\infty, s) \approx \left[1+\frac{2}{D}\Bigl(\langle E \rangle_\alpha s\Bigr)^\alpha\right]^{-D/2\alpha},
    \label{eq:mittagleffler}
\end{align}
is the simplest approximation of the steady-state solution of \eref{eq:recursive_dissipation}.
The GML distribution naturally reduces to the Maxwell distribution at $\alpha\rightarrow 1$.

\begin{figure}
    \includegraphics[width=7.5cm]{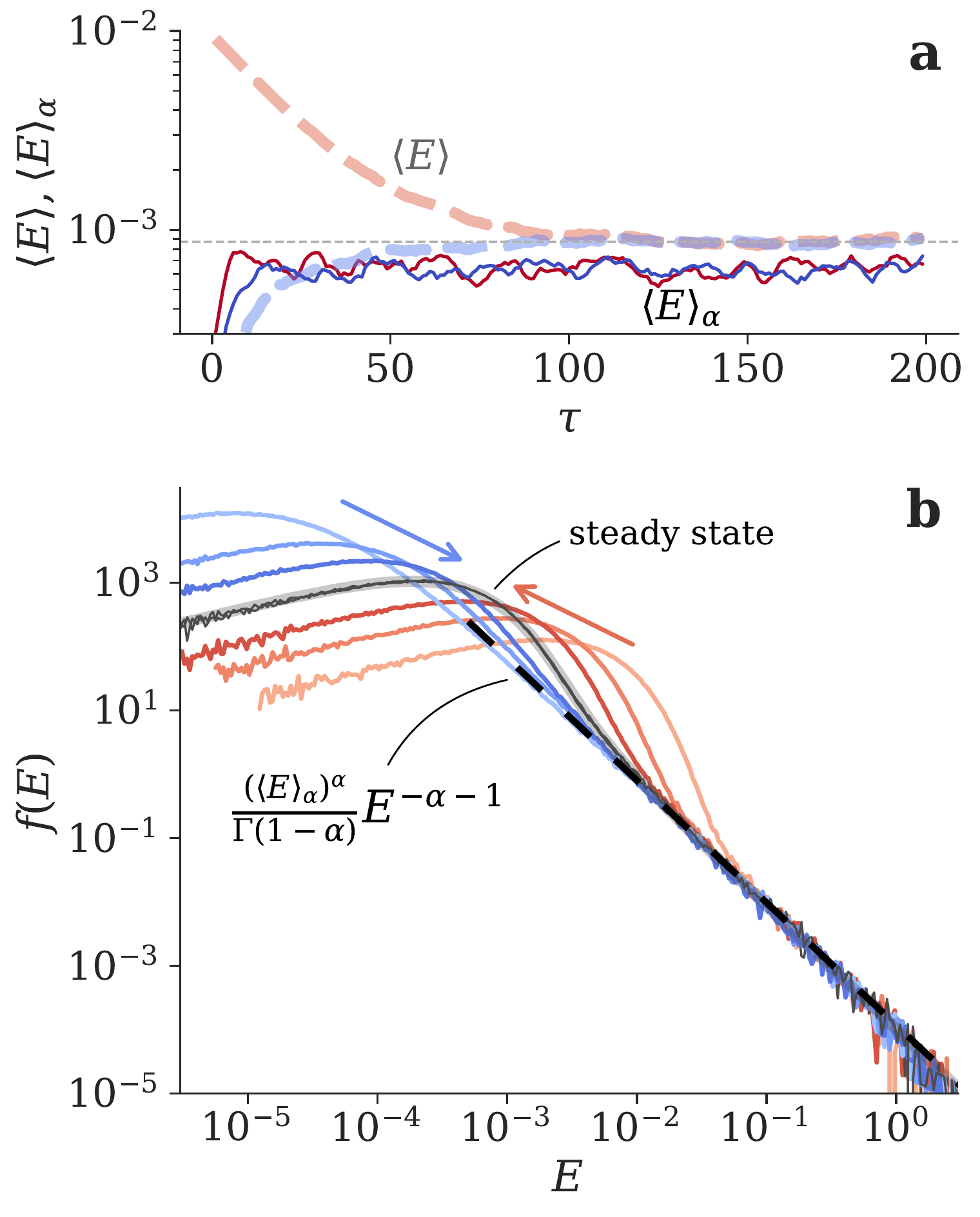}
    \caption{%
    Temporal evolution of inelastic particles with Maxwell interaction and extreme heating.
    Two simulation runs started from different initial distributions are shown, where red curves for the run with the hotter initial condition and blue curves for the colder condition.
    (a) Temporal evolutions of $\langle E\rangle$ and $\langle E\rangle_\alpha$.
    $\langle E\rangle$ decays toward the steady-state value while $\langle E\rangle_\alpha$ stays constant from the beginning.
    (b) %
    Energy distributions for the two simulation runs at several time slices.
    The distributions evolve along the arrow direction and arrive at the same steady-state distribution (black curves).
    Gray bold curve shows the best fit by the GML distribution, with the optimum value of $\alpha=0.92$. 
    Dashed straight line shows the power-law $(\langle E \rangle_\alpha)^\alpha E^{-\alpha-1}/\Gamma(1-\alpha)$.
    }
    \label{fig:granular_temporal}
\end{figure}

As a demonstration, we carry out a Monte-Carlo simulation for a spatially uniform ensemble of inelastic particles having the Maxwell interaction, as done by Ben-Naim et al.~\cite{Ben-Naim2005-rd,Ben-Naim2005-uz} [for the comparison with more realistic simulations and experimental observations, see Supplemental Material~\cite{Note1} and an accompanying paper~\cite{Fujii2022_PRE}].
At every step of the simulation, we randomly choose the colliding pairs of particles and compute its scattering and energy dissipation based on the collision geometry.
We consider inelastic collision, where the relative velocity along the collision normal is reduced by the factor of $1-r$ with elasticity $r$~\cite{Brilliantov2004-dj,Aranson2006-qf}.
As an energy injection process, we choose a particle randomly at a certain rate and replace their velocity to Maxwellian with temperature 1.
We keep the energy injection rate constant and continue the simulation until the system reaches the steady state.

\Fref{fig:granular_temporal} shows the simulation results for $r=0.9$.
Dashed lines in \fref{fig:granular_temporal}~(a) show the temporal evolution of the mean energy $\langle E \rangle$ for the system.
Two different colored curves show the values of $\langle E \rangle$ for two simulation runs started from different initial distributions.
As this is a nonthermal system, the mean energies are not conserved and evolves in time toward the steady-state value.
\Fref{fig:granular_temporal}~(b) shows the energy distributions at several time slices during the temporal evolution of the two simulation rns (blueish curves are from the simulation with the lower initial energy, and reddish curves from that with the higher initial energy).
A black curve in the figure is the steady-state distribution.
All the distributions have the power-law tail and its amplitude and slope are constant in time.

The thin curves in \fref{fig:granular_temporal}~(a) show the temporal evolution of $\langle E \rangle_\alpha$ obtained from the distribution. 
They stay unchanged during the evolution, which is in contrast with the decay of $\langle E \rangle$. 
This is consistent with our above argument, where $\langle E \rangle_\alpha$ is a conserved quantity.
Note that as there is an energy cut-off at $E \approx 1$ in our system, the direct computation from \eref{eq:fractional_mean} is not feasible. 
Instead, it is estimated from the distribution tail, based on the fact that the power-law tail is written with the fractional mean, 
$(\langle E \rangle_\alpha)^\alpha E^{-\alpha-1} / \Gamma(1-\alpha)$, 
where $\alpha=0.92$ is taken from the best-fit by the GML distribution (see below).

\begin{figure*}
    \includegraphics[width=17cm]{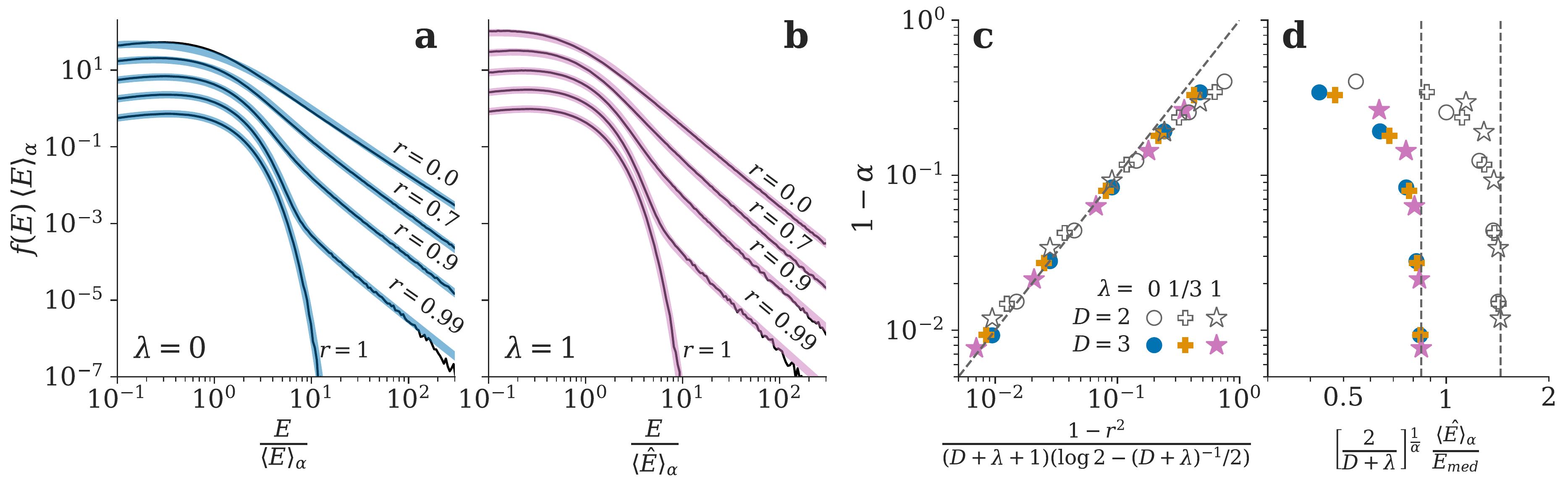}
    \caption{%
        Steady state distributions of isotropic inelastic gases by the Monte-Carlo simulations.
        Thin curves: the simulation results for (a) the gases with Maxwell interaction and (b) for the hard spheres.
        Results with several values of the inelasticity $r$ are shown, with an appropriate vertical offset for the sake of clarity.
        Thick curves: the best fit by the GML distribution.
        (c) The optimum values of $\alpha$. The horizontal position of the markers are computed from \eref{eq:alpha}.
        (d) The values of $\langle \hat{E} \rangle_\alpha$ scaled by the median energy $E_{med}$.
        The vertical dashed lines are the corresponding values of $\langle E \rangle/E_{med}$ for the Maxwell distribution.
    }
    \label{fig:granular}
\end{figure*}

The steady-state distribution has a power-law tail in the high-energy region, as pointed out in the original works~\cite{Ben-Naim2005-rd,Ben-Naim2005-uz}.
The low-energy region is similar to the Maxwellian (see also \fref{fig:granular}~(a) later).
The bold curve in the figure shows the best fit by the GML distribution \eref{eq:mittagleffler}.
The GML distribution well reproduces the simulated result.

\Fref{fig:granular}~(a) shows the steady-state distribution simulated with different values of $r$.
The distribution with $r=1$ (elastic limit, with no heating) falls exponentially in the high-energy region, while with $r < 1$ the distribution has a power-law tail.
We find bigger tails in the distribution with larger energy dissipation, i.e., the smaller values of $r$.

The bold curves in \fref{fig:granular}~(a) show the best-fit by the GML distribution.
The GML distribution well represents the simulated energy distributions, particularly those under the small dissipation.
The best-fit values of $\alpha$ is shown in \fref{fig:granular}~(c) by filled circles as a function of $r$.
With the smaller energy dissipation, $\alpha$ closer to 1 is obtained.

The value of $\alpha$ may be analytically computed from \eref{eq:alpha} by using diffuse kernel \eref{eq:diffuse}. 
From the averaged energy loss of one inelastic collision $1-e^{-\Delta}\approx (1-r^2)/2(D+1)$ and $\xi=1/2$, we obtain
\begin{align}
    1- \alpha \approx \frac{1-r^2}{
        \left(D + 1\right)
        \left(\log 2 - \frac{1}{2D}\right)
    },
    \label{eq:alpha_granular}
\end{align}
Here, we assume $0 \leq \Delta \ll 1$.
The dotted diagonal line in \fref{fig:granular}~(c) shows \eref{eq:alpha_granular}.
The filled circles are well aligned on this prediction, particularly when $1-r \ll 1$.

Filled circles in \fref{fig:granular}~(d) shows the value of $\langle E \rangle_\alpha$ scaled by the median energy $E_{med}$.
This approaches to the corresponding value of the Maxwell distribution in the 3-dimensional space, $\langle E \rangle / E_{med}\approx 0.56$ (vertical dashed line) with $\alpha \rightarrow 1$.

The above discussion can be approximately extended to particles having other inter-particle interactions.
For example, the collision rate of hard spheres is proportional to $E^{\lambda/2}$ with $\lambda=1$ while neutral atomic gases show Van-der-Waals interaction, where $\lambda=1/3$~\cite{Massey1934, Flannery2006}.
For such systems, we may consider the weighted distribution, $\hat{f}(E) = E^{\lambda/2} f(E) / Z$, with the normalization constant $Z$.
Based on an approximation $(E_1 + E_2)^{\lambda/2} \approx (E_1 E_2 / \langle E \rangle_\alpha)^{\lambda/2}$, which is valid if $|\lambda| \ll D$, this weighting approximately represents the energy dependence of the collision rate.
Although this weighting changes the statistical weight of the $D$-dimensional space from $\propto E^{D/2-1}$ to $\propto E^{(D + \lambda)/2 - 1}$, the Laplace transform of its weighted distribution at the steady state is approximated by the GML distribution, $\mathcal{L}_{\hat{f}}(s)=[1+(D + \lambda)(\langle E \rangle_\alpha s)^\alpha / 2]^{-(D+\lambda)/2\alpha}$.
In this case, the quantity approximately conserved during the temporal evolution is obtained by replacing $f(E)$ by $\hat{f}(E)$ in \eref{eq:fractional_mean},
$\hat{\langle E \rangle}_\alpha \equiv
    \left[\lim_{E\rightarrow \infty}
        \int_0^E (E-x)^{\alpha-1} x \hat{f}(x) dx / \Gamma(\alpha)
        \right]^{1/\alpha}$.
Note that although $\langle\hat{E}\rangle_\alpha$ is not reduced exactly to $\langle E \rangle$ at $\alpha=1$ because of the approximation to take $\lambda\neq 0$ interaction into account, it gives a good approximation as long as $|\lambda|\ll D$.

Similar Monte-Carlo simulations are carried out with $\lambda=1$ and $1/3$.
In these runs, the relative velocity among particles are taken into account in choosing colliding pairs of particles.
Thin curves in \fref{fig:granular}~(b) show the steady-state energy distributions for hard spheres in 3-dimensional space for various values of $r$.
The GML distribution (bold curves) well represents the steady-state distributions also for $\lambda = 1$ cases.
\Fref{fig:granular}~(c) shows the optimum values of $\alpha$, for $\lambda=0, 1/3$ and 1, and $D=2$ and 3 cases.
The horizontal positions of the markers are computed from \eref{eq:alpha_granular} but with $D$ replaced by $D + \lambda$. 
All the results are well aligned on the diagonal line, indicating the consistency with the above discussion.

\Fref{fig:granular}~(d) shows the value of $\langle E\rangle_\alpha$ scaled by the median energy.
Also for these $\lambda\neq 0$ simulations, $\langle E\rangle_\alpha / E_{med}$ converges to the values in the thermal system (vertical dotted lines), showing the smooth transition from the nonthermal to thermal systems.

In this Letter, I pointed out that the fractional mean energy is conserved during a temporal evolution of dissipative gases. 
This conservation law is equivalent with the time-invariant power-law tails in the energy distribution.
The distribution approaches to the Maxwell distribution and the fractional mean energy converges to the standard mean energy, as we tune the system close to the no-dissipation limit.
It is also pointed out that the steady-state distribution is well approximated by the GML distribution, an application of which to plasma-physics field is separately reported in Ref.\cite{Fujii2022_PRE}.

The power-law tail in the energy distribution is ubiquitous in many dissipative systems, such as cosmic rays accelerated in shock fronts~\cite{Bell1978-wg}, earthquakes~\cite{Gutenberg1944-wb}, and fluid turbulence~\cite{Kolmogorov1991-uo,Hasegawa1977-oz}.
Although we focused only on gaseous systems in this work, a similar conservation law is expected for other systems.

The establishment of the thermodynamics for nonthermal systems,
particularly the nonthermal equivalence of the first- and second-laws, is one of long-standing open questions in physics. 
The conservation of the fractional mean energy may be interpreted as the first-law equivalence.
Although several generalizations of the entropy have been proposed as the second law for nonthermal systems~\cite{Renyi2007-zw,Tsallis1988-ef,Landsberg1998-lf}, it is found that none of them is consistent with the system we considered here as well as our conservation law.
The search of an entropy form for the second law is in the scope of future studies.

\begin{acknowledgments}
    This work was supported by the U.S. D.O.E contract DE-AC05-00OR22725.
    An anonimous person with the username \textit{vitamin d}, who gave me an essential suggestion in \url{https://mathoverflow.net/questions/401835/} is also appreciated.
    Also, the author thanks fruitful comments from Dr. Maeyama (Nagoya University), Dr. Shiba (University of Tokyo), and Dr. Del-Castillo-Negrete (ORNL).
\end{acknowledgments}

\bibliography{refs}

\input{supplemental/supplemental.tex}
    
\end{document}

%% file: supplemental/supplemental.tex
\pagebreak

\newcommand{\ManuscriptTitleSupp}{\ManuscriptTitle}

\newcommand{\EqLaplaceNodissipation}{eq:laplace_nodissipation}
\newcommand{\EqRecursive}{eq:recursive} 
\newcommand{\EqLaplaceNoDissipation}{eq:laplace_nodissipation}
\newcommand{\EqDiffuse}{eq:diffuse} 
\newcommand{\EqLimitting}{eq:limitting}
\newcommand{\EqAlpha}{eq:alpha} 
\newcommand{\EqConstantAlpha}{eq:constant_alpha}
\newcommand{\EqFractionalMean}{eq:fractional_mean}
\newcommand{\EqMittagLeffler}{eq:mittagleffler}

\widetext
\begin{center}
\textbf{\large \ManuscriptTitleSupp}
\end{center}
\setcounter{equation}{0}
\setcounter{figure}{0}
\setcounter{table}{0}
\makeatletter

\renewcommand{\theequation}{S\arabic{equation}}
\renewcommand{\thefigure}{S\arabic{figure}}
\renewcommand{\thetable}{S\arabic{table}}

\section{Derivation of the Conservation Law and GML distributions}
In this section, the detailed derivations of some equalities are presented.

\subsection{Detailed derivation of $d\alpha / d\tau = 0$ and $dc / d\tau = 0$}
Let us consider the small-$|s|$ limit of $\mathcal{L}_f(\tau, s)$.
The normalization condition gives $\mathcal{L}_f(\tau, 0) = 1$. 
From the second two smallest orders, $\mathcal{L}_f(\tau, s)$ can be written as $\mathcal{L}_f(s) \approx 1 - c(\tau) s^{\alpha(\tau)}$ at small-$|s|$ region.
Here, $c(\tau) > 0$ should be satisfied according to a property of the Laplace transform.
By substituting it into \eref{\EqRecursive}, we obtain
\begin{align}
    \notag
    -\frac{\partial}{\partial \tau} c(\tau) s^{\alpha(\tau)}
    \notag
    & = - (1 - c s^\alpha) + 
    (1-\xi)\int (1 - c x^\alpha s^\alpha - c y^\alpha s^\alpha)p(x, y)dxdy + \xi (1 - c e^{-\alpha \Delta}s^\alpha)\\
    &= c s^\alpha\left[
        1-\left\{
            (1-\xi)\int(x^\alpha+y^\alpha)p(x,y)dxdy + \xi e^{-\alpha \Delta}
        \right\}
    \right],
\end{align}
which leads
\begin{align}
    \frac{\frac{\partial}{\partial \tau}c(\tau)}{c(\tau)} + 
    \frac{1}{s} \frac{\partial}{\partial \tau}\alpha(\tau)
    = 
    (1-\xi)\int(x^\alpha+y^\alpha)p(x,y)dxdy + \xi e^{-\alpha \Delta} - 1.
    \label{sup:eq:time_evolution}
\end{align}
By equating the terms for $s^0$ and $s^{-1}$ in both the sides, we obtain
\begin{align}
    \frac{\frac{\partial}{\partial \tau}c(\tau)}{c(\tau)} &= 
    (1-\xi)\int(x^\alpha+y^\alpha)p(x,y)dxdy + \xi e^{-\alpha \Delta} - 1,\\
    \label{sup:eq:dcdt}
    \frac{\partial}{\partial \tau}\alpha(\tau) &= 0,
\end{align}
respectively.

Here, we define a new quantity $\alpha_0$, which satisfies \eref{\EqAlpha} if substituted as $\alpha$.
First, let us consider the case of $\alpha > \alpha_0$.
Note that the coefficient for the order of $s^{\alpha_0}$ is zero according to our definition.
Then, we get $\frac{\partial}{\partial \tau}c(\tau)/ c(\tau) < 0$.
In this case, $c(\tau)$ exponentially decays and after long enough time the contribution of $s^\alpha$ becomes negligible.
We can repeat the same discussion for the next order to $\alpha$.
Eventually we find that all the terms in any order decays to zero, i.e., the steady state has the zero kinetic energy.
This corresponds to the case with no heating source to the system, which is against our assumption that the system has a nontrivial steady state.
Note that this case corresponds to $\langle E \rangle_\alpha = 0$.

Secondly, let us consider the case of $\alpha < \alpha_0$.
Then, we get $\frac{\partial}{\partial \tau}c(\tau)/ c(\tau) > 0$.
$c(\tau)$ diverges exponentially and does not reach the steady-state.
This is again contradictory to our assumption.
Note that this situation corresponds to $\langle E \rangle_\alpha = \infty$.

Therefore, the system arriving at a nontrivial steady state should satisfy \eref{\EqAlpha}. 
Note that in realistic systems, the energy of the heat source is finite and thus with the nonzero power input, this relation is always satisfied.

\subsection{Derivation of \eref{\EqFractionalMean}}

Let us define a polynomial function $g(x) = x^{\alpha-1} / \Gamma(\alpha)$ so that the Laplace transform of $\mathcal{L}_g = s^{-\alpha}$.
Recall that the convolution of $g(x)$ and another function $h(x)$, i.e., $g(x) * h(x) \equiv \int_0^x g(x-t) h(t) dt$ is equivalent with the product of their Laplace transforms.
We obtain,
\begin{align}
    \lim_{|s|\rightarrow 0}s^{1-\alpha} \frac{\partial \mathcal{L}_f(\tau, s)}{\partial s} 
    & =
    \lim_{|s|\rightarrow 0} s \mathcal{L}_{g(x) * (x f(x))} \\
    & = \lim_{|s|\rightarrow 0} \int_0^\infty dE\, s e^{-sE} \int_0^E dx\, g(E-x) x f(x)\\
    & = \lim_{|s|\rightarrow 0} \frac{1}{\Gamma(\alpha)}\int_0^\infty dE\, e^{-sE} \frac{d}{dE} 
    \left[
        \int_0^E (E-x)^{\alpha-1} x f(x) dx
    \right].
\end{align}
Because $(d/dE) \left[
    \int_0^E (E-x)^{\alpha-1} x f(x) dx
\right]$ is nonnegative for all $E\geq 0$ and integrable, we can exchange the limit and the integration according to the dominated convergence theorem.
This yields \eref{\EqFractionalMean}.

\subsection{Derivation of \eref{\EqMittagLeffler}}

In order to obtain an asymptotic form of $\mathcal{L}_f(\infty, s)$ at the steady state, we consider \eref{\EqLaplaceNoDissipation} with large and real $s$.
Since $\mathcal{L}_f(\infty, s)$ is the monotonically decreasing nonnegative function of $s$, the dominant contribution to the integrand in \eref{\EqMittagLeffler} comes from the small-$x$ and $y$ region so that $xs, ys \lesssim 1$.
Let $p(x,y) \approx c_p x^a y^b$ be the smallest order approximation of $p(x,y)$.
Also, from the consideration of the small-$|s|$ limit of $\mathcal{L}_f$, we may approximate $\mathcal{L}_f(\infty, s) \approx [1 + \nu^{-1} (\langle E \rangle_\alpha s)^\alpha]^{-\nu}$ where $\nu$ is an unknown parameter.
By substituting them, the integral in \eref{\EqMittagLeffler} can be written as
\begin{align}
    \notag
    \int\, \mathcal{L}_f(\infty, xs) \mathcal{L}_f(\infty, ys)\, p(x, y) \,dx\,dy
    &\approx
    c_p \int_0^\infty[1 + \nu^{-1} (\langle E \rangle_\alpha xs)^\alpha]^{-\nu} x^a dx 
    \int_0^\infty[1 + \nu^{-1} (\langle E \rangle_\alpha ys)^\alpha]^{-\nu} y^b dy\\
    &=s^{-(a+b+2)} 
    c_p \left(
        \frac{\nu^{1/\alpha}}{\langle E \rangle_\alpha}
    \right)^{a+b+2}
    \alpha^{-2}
    B\left(\frac{a}{\alpha}+1, \nu - \frac{a}{\alpha} - 1\right)
    B\left(\frac{b}{\alpha}+1, \nu - \frac{b}{\alpha} - 1\right).
    \label{sup:eq:large_s}
\end{align}
As $a$ and $b$ do not depend on $\alpha$, we obtain $a + b + 2 = D/2$ by considering the thermal system. Thus, this integration should be proportional to $s^{-D/2}$ with large $s$, independent of $\nu$ and $\alpha$.
In order to match this dependence to that of the rest of the terms, $\mathcal{L}_f(\infty, s)$ should be proportional to $s^{-D/2}$ at the large-$s$ region.
At the thermal system, $\mathcal{L}_f(\tau, s) \propto 2D^{-1}(\langle E \rangle_\alpha s)^{-D/2}$.
As the rest of the terms in \eref{sup:eq:large_s} only depend on $\alpha$ in the first order.
Thus, with $\alpha \lesssim 1$, $\mathcal{L}_f(\tau, s)$ is written as $\propto [2D^{-1} + \mathcal{O}(1-\alpha)](\langle E \rangle_\alpha s)^{-D/2}$.

\section{Numerical evaluation of GML distribution}

In the main text, the numerical fit by the GML distribution is carried out.
Although the GML distribution has no analytical forms except for few special cases, an efficient numerical computation method has been proposed~\cite{haubold_mittag-leffler_2011,barabesi_new_2016,korolev_mixture_2020},
\begin{align}
    &f_{\operatorname{GML}}(E|\alpha, D, \langle E \rangle_\alpha) = 
    \frac{1}{\pi \langle E \rangle_\alpha} 
    \left(\frac{D}{2}\right)^{\frac{1}{\alpha}}
    \int_0^\infty \frac{
        \exp\left(-y\left(\frac{D}{2}\right)^{\frac{1}{\alpha}}\frac{E}{\langle E \rangle_\alpha}\right)
        \sin\left(\pi\frac{D}{2} F_\alpha(y)\right)
    }{
       \left(y^{2\alpha} + 2 y^\alpha \cos(\pi \alpha) + 1\right)^{D/4\alpha}
    }
    dy,
    \label{sup:eq:gml_mixture}
\end{align}
where $f_{\operatorname{GML}}(E|\alpha, D, \langle E \rangle_\alpha) = \mathcal{L}^{-1}\left[(1+2(\langle E \rangle_\alpha s)^\alpha / D)^{-D/2\alpha}\right]$.
Here $F_\alpha(y)$ is defined as follows,
\begin{align}
    F_\alpha(y) = 1 - \frac{1}{\pi\alpha}\cot^{-1}\left(
        \cot(\pi\alpha) + \frac{y^\alpha}{\sin(\pi\alpha)}
    \right).
\end{align}
In this work, the values of the GML distribution is evaluated by integrating \eref{sup:eq:gml_mixture} numerically.

\begin{figure}
    \includegraphics[width=7.5cm]{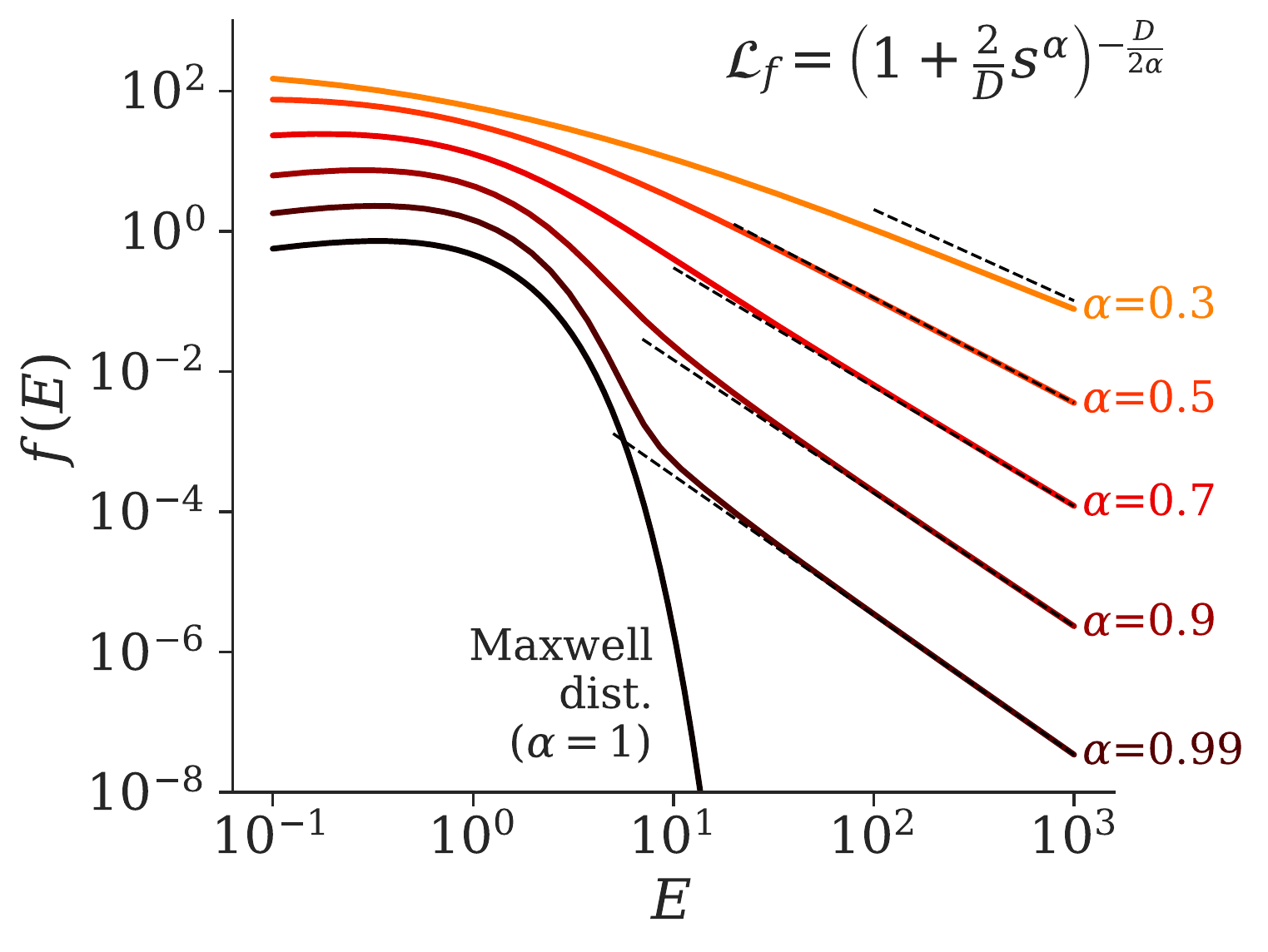}
    \caption{%
    The GML distribution with several values of $\alpha$ with $\langle E \rangle_\alpha = 1$. The power-law tail $E^{-\alpha-1}/\Gamma(1-\alpha)$ is shown by dotted lines.
    An appropriate offsets are introduced for the sake of clarity.
    }
    \label{sup:fig:mittagleffler}
\end{figure}

\Fref{sup:fig:mittagleffler} shows the GML distribution for several values of $\alpha$. 
As we see from the small-$|s|$ dependence, it has a power-law tail, 
$(\langle E\rangle_\alpha)^{\alpha} E^{-\alpha-1}/\Gamma(1-\alpha)$.
The dotted lines in the figure are this power-law function.
The GML distribution approaches to this power-law tail in the large-$E$ region.
The power-law tail becomes smaller as $\alpha$ approaches to 1, and at $\alpha = 1$, the power-law tail disappears and the distribution reduces to the Maxwell distribution.

\section{Probabilistic Representation of Elastic Collisions}
In the main text, the energy change by an elastic collision is modeled by a probabilistic form \eref{\EqRecursive}.
In this section, the details of the assumptions, necessary conditions, as well as the actual form for some particular cases are presented.

\subsection{Necessary condition for a valid $p(x,y)$}

The form of $p(x, y)$ in \eref{\EqRecursive} should depend on the inter-particle interaction. 
Although in the next subsection a particular case (hard-sphere collision) will be discussed, here let us consider the necessary condition for a valid $p(x,y)$.

First, as we consider the elastic collision, the sum of the kinetic energies should be conserved, i.e., $E_1 + E_2 = E_1' + E_2'$, where $E_2'$ is the post-collision energy of particle 2.
The similar relation for particle 2 is
\begin{align}
    E_2' \leftarrow (1-y) E_2 + (1-x) E_1.
\end{align}
The exchange of particles 1 and 2 gives the following symmetry condition,
\begin{align}
    p(x, y) = p(1-y, 1-x),
\end{align}
which directly leads
\begin{align}
    \int_0^1 (x + y) p(x, y) dx\,dy = \int_0^1 p(x, y) dx\,dy = 1.
\end{align}
Find that \eref{\EqAlpha} reduces to the above equation when substituting $\alpha=1$ and $\xi = 0$.

Second, the reverse reaction should have the same probability, i.e., $p(x, y)$ should satisfy the detailed balance.
Let us consider the two variables $z \equiv E_1/(E_1+E_2)$ and $z' \equiv E_1'/(E_1'+E_2')$.
The conditional probability distribution of $z'$ with given $z$ is
\begin{align}
    p(z'|z) = 
    \frac{1}{z} \int_{\max(0, \frac{z'-z}{1-z})}^{\min(1, \frac{z'}{1-z})}
    p\left(
        \frac{z'}{z} - \frac{1-z}{z}y, \,y
    \right)
    dy.
\end{align}
The detailed balance can be written as
\begin{align}
    p(z'|z) B\left(z \middle| \frac{D}{2}, \frac{D}{2}\right)
    = p(z|z') B\left(z' \middle| \frac{D}{2}, \frac{D}{2}\right).
    \label{sup:eq:detailed_balance}
\end{align}
$p(x,y)$ should satisfy \eref{sup:eq:detailed_balance} for any pair of $z$ and $z'$.
Note that the beta distribution $B(z|D/2, D/2)$ represents the statistical weight of $z$ in the $D$-dimensional space.
The diffuse collision \eref{\EqDiffuse} and the $p$-$q$ model, which we will discuss below, satisfies this detailed balance relation.

\begin{figure*}[b]
    \centering
    \includegraphics[width=10cm]{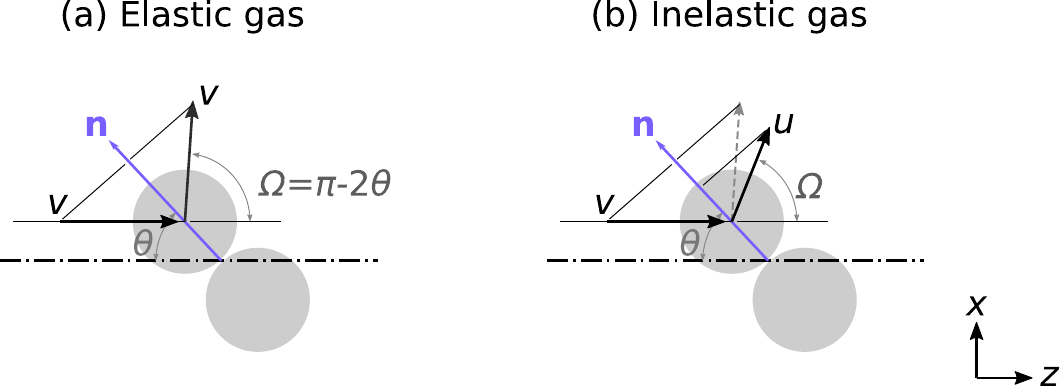}
    \caption{
        Schematic illustration of the two-body collision in the center-of-mass frame.
        (a) Elastic collision and (b)inelastic collision.
    }
    \label{sup:fig:collision}
\end{figure*}

\subsection{Exact Description of Elastic Collision of Hard Spheres}
Let us consider an elastic collision among two hard spheres having mass 1 (\fref{sup:fig:collision}~(a)).
Before the collision, two hard spheres have velocities $\mathbf{v}_1$ and $\mathbf{v}_2$. 
The center-of-mass (CM) velocity and their relative velocity is 
\begin{align}
    \mathbf{V}_{CM} &= \frac{\mathbf{v}_1 + \mathbf{v}_2}{2},\\
    \mathbf{v} &= \frac{\mathbf{v}_1 - \mathbf{v}_2}{2},
\end{align}
respectively.
Let $\Omega$ be the scattering angle in the CM frame.
The post-collision velocity of the particle 1, $\mathbf{v}'_1$, has the following relation with the pre-collision velocities,
\begin{align}
    v \equiv |\mathbf{v}'_1 - \mathbf{V}_{CM}| = |\mathbf{v}_1 - \mathbf{V}_{CM}|,\\
    (\mathbf{v}'_1 - \mathbf{V}_{CM}) \cdot (\mathbf{v}_1 - \mathbf{V}_{CM}) = v \cos\Omega.
\end{align}
After a simple equating, we obtain the following relation between the pre-collision energies $E_1, E_2$ and post-collision energy $E'_1$,
\begin{align}
    E'_1 = \frac{1}{2}\left[
        (E_1 + E_2) - (E_2 - E_1)\cos\Omega - 2\sqrt{E_1 E_2} r \sin\Psi \sin\Omega
    \right],
\end{align}
where, $\Psi$ is the angle between $\mathbf{v}_{1}$ and $\mathbf{v}_{2}$, $r$ is the cosine angle between $\mathbf{V}_{CM}$ and the plane spanned by $\mathbf{v}_{rel}$ and $\mathbf{v}'_{rel}$.
$\Omega = \pi - 2 \theta$, $\Psi$, and $r$ are independent of each other and they follow
\begin{align}
    \cos^2\theta &\sim B\left(\frac{1}{2}, \frac{D-1}{2}\right),\\
    \cos^2\Psi &\sim B\left(\frac{1}{2}, \frac{D-1}{2}\right),\\
    r^2 &\sim B\left(\frac{1}{2}, \frac{D-2}{2}\right).
    \label{sup:eq:exact_collision}
\end{align}
   
\Fref{sup:fig:collision_kernel}~(a) shows the probability distribution of $p(E'_1)$ for $D=3$ case.
For comparison, the distribution by \eref{\EqDiffuse} is shown in \fref{sup:fig:collision_kernel}~(b), which has been employed to study gas kinetics for a long time.
Despite of a small correlation in \fref{sup:fig:collision_kernel}~(a), the overall distribution is similar to the diffuse model.

\subsection{The $p$-$q$ Model for Elastic Collision}
\begin{figure*}
    \includegraphics[width=17cm]{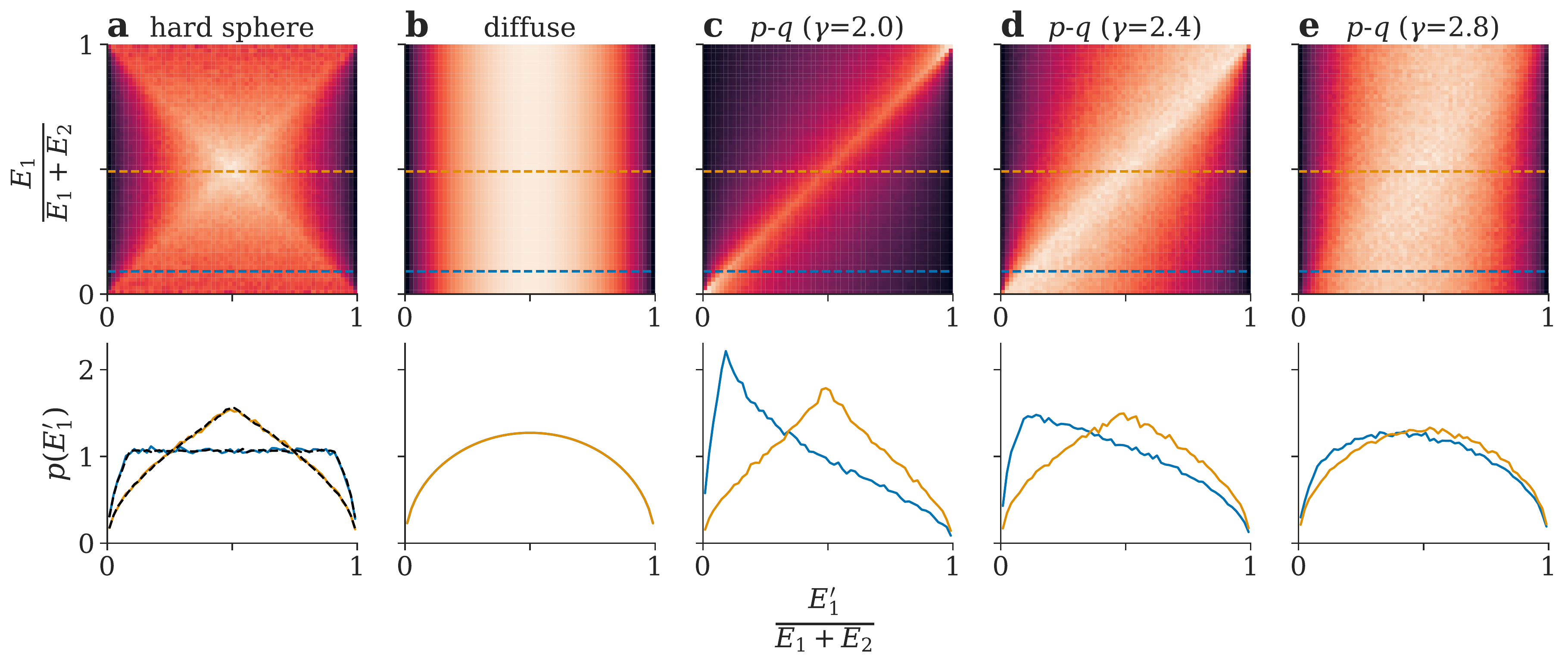}
    \caption{%
        Probability distribution of the pre-collision energy $E_1$ and the post-collision energy $E_1'$.
        The lower panel shows the crosssections at the two $E_1'$ values (at the position of the horizon lines in the upper panel).
        (a) The exact distribution for the hard-sphere collision, (b) the diffuse collision model \eref{\EqDiffuse}, (c)-(e) the distribution by $p$-$q$ model with some values of $\gamma$.
        The dotted curves in the lower panel of (a) shows the best-fit of the exact distribution by \eref{sup:eq:pq_superposition}.
    }
    \label{sup:fig:collision_kernel}
\end{figure*}

To capture the correlation found in the exact $p(E'_1)$, so-called $p$-$q$ model has been proposed~\cite{Futcher1983-sf,Futcher1980-ey}.
This model is equivalent to the following probabilistic process
\begin{align}
    E'_1 = (1-a) E_1 + (a E_1 + b E_2) c,
    \label{sup:eq:pq-model}
\end{align}
where $a$, $b$, and $c$ are the independent random variables, following
\begin{align}
    a &\sim B\left(a \middle| \frac{\gamma}{2}, \frac{D-\gamma}{2}\right),\\
    b &\sim B\left(b \middle| \frac{\gamma}{2}, \frac{D-\gamma}{2}\right),\\
    c &\sim B\left(c \middle| \frac{\gamma}{2}, \frac{\gamma}{2}\right),
\end{align}
where $0 \leq \gamma \leq D$ is a constant that controls the strength of the correlation.
\Frefs{sup:fig:collision_kernel}~(c), (d), and (e) shows the distribution of $p(E'_1)$ for several values of $\gamma$.
Depending on the value of $\gamma$, $p(E'_1)$ changes from a strong memory collision (with small $\gamma$) to a nearly-diffuse collision (with large $\gamma$).

\Eref{sup:eq:pq-model} has the form of \eref{\EqRecursive}, where
\begin{align}
    p(x,y|\gamma) = 
    \frac{(1-x)^{\gamma/2-1}y^{\gamma/2-1}}{
        B\left(\frac{\gamma}{2}, \frac{\gamma}{2}\right)
        \left(B\left(\frac{\gamma}{2}, \frac{D-\gamma}{2}\right)\right)^2}
    \int_y^x
        c^{(\gamma - D)/2-2}(1-c)^{(\gamma-D)/2-2}(x-c)^{(D-\gamma)/2-1}(c-y)^{(D-\gamma)/2-1}
    dc
\end{align}
It can be easily shown that this $p(x,y | \gamma)$ with any value of $\gamma$ is a valid probability distribution that leads the Maxwell distribution at the steady state when used in \eref{\EqLaplaceNoDissipation}.
Similarly, $p(y,x|\gamma)$ is also a valid distribution.
Furthermore, the mixture of $p(x,y | \gamma)$ and $p(y,x|\gamma)$ for different values of $\gamma$ is also valid, which is the linear superposition of $p(x,y | \gamma)$ and $p(y,x | \gamma)$ with arbitrary weight distributions $p_1(\gamma)$ and $p_2(\gamma)$,
\begin{align}
    p(x,y)
    =& \eta\int_0^{D} p(x,y|\gamma) p_1(\gamma) d\gamma
    +(1-\eta)\int_0^{D} p(y,x|\gamma) p_2(\gamma) d\gamma.
    \label{sup:eq:pq_superposition}
\end{align}
Here, $0\leq\eta\leq 1$ is the relative weight of the two terms.
Dotted curves in the lower panel of \fref{sup:fig:collision_kernel}~(a) shows the best fit of the exact kernel $p(x,y)$ (\eref{sup:eq:exact_collision}) by \eref{sup:eq:pq_superposition}.
This perfectly represents the exact solution.
Because of the flexibility in \eref{sup:eq:pq_superposition}, most of the realistic collision can be represented by \eref{\EqRecursive}.

\section{Inelastic Collision}

One of the standard models for an inelastic collision is to adopt an inelasticity for collision velocity~\cite{Brilliantov2004-dj,Aranson2006-qf}.
When two particles undergo an inelastic collision, the scattering angle $\Omega$ depends on the restitution coefficient $0 \leq r \leq 1$ (\fref{sup:fig:collision}~(b)).
Because of the inelasticity, the momentum normal to the collision direction (vector $\mathbf{n}$ in the figure) changes~\cite{villani_mathematics_2006},
\begin{align}
    v \cos \theta = r u \cos \theta,
    \label{sup:eq:pdelta}
\end{align}
with the post-collision velocity $u$ in the CM frame.
The momentum perpendicular to $\mathbf{n}$ is conserved.
$u$ and the scattering angle $\Omega$ can be written as
\begin{align}
    \label{sup:eq:velocity_loss}
    u &= v \sqrt{\sin^2 \theta + r^2 \cos^2 \theta},\\
    \Omega &= \cos^{-1}\left(-\frac{r\cos\theta}{\sqrt{\sin^2 \theta + r^2\cos^2 \theta}}\right) - \theta.
\end{align}

In this work, we use the hard-sphere cross section for the inelastic gas.
By averaging \eref{sup:eq:velocity_loss}, we obtain $ (v^2 - \langle u^2 \rangle)/2 \approx (1-r^2)/(D+1)$.
Since in an isotropic system, the kinetic energy should be shared equally by the kinetic energy in the CM frame and that of the center of mass, i.e., $\langle v^2 \rangle = \langle V_\mathrm{CM}^2 \rangle$,
the average of the fractional energy loss per one collision is $(1-r^2)/2(D+1)$.

\section{Demonstration with a Direct Molecular Simulation}

\begin{figure}[b]
    \includegraphics[width=7.5cm]{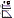}
    \caption{%
    (a) A schematic illustration of our molecular-dynamics simulation, where atoms are confined by inelastic walls.
    (b) The temporal evolutions of $\langle E \rangle$ and $\langle E \rangle_\alpha$ for two simulation runs, and
    (c) the energy distributions at several time slices, as similar to \fref{fig:granular_temporal}.
    }
    \label{sup:fig:md_temporal}
\end{figure}

In the main text, the Monte-Carlo simulations for the inelastic gases are presented.
In these simulations, a spatially uniform and isotropic gas, as well as the molecular chaos are assumed, i.e., the spatial correlation after collisions is neglected.

In order to see the validity of the main argument for more realistic situations, here I show the direct molecular dynamics simulation for atomic gas surrounded by cold walls.
As shown in \fref{sup:fig:md_temporal}~(a), $N$ atoms with mass $m$ in a cubic box with one side of $L$ are considered.
These atoms interact according to the inter-atomic potential $V(r) = (r/r_0)^{-6}$, where $r$ is the inter-atomic distance.
Thus, the atom-atom collision is elastic.
We assume that the walls have infinitely large degrees-of-freedom and have much lower temperature than the atomic gas.
A collision with such a wall can be approximated by an inelastic collision~\cite{Ito1985-px}.
We simulate such a wall collision by an inelastic coefficient $r$, where the atomic velocity perpendicular to the wall changes $v_{\perp} \rightarrow -rv_{\perp}$.
The box has an opening with the area of $a$. 
If an atom goes out of the box through this opening, another atom having the temperature $T_0$ is injected into the box.
At the steady state, this energy injection will be balanced with the energy dissipation by the wall collision. 

This system mimics the neutral gas behavior in plasmas.
Neutral atoms, particularly radical atoms, gain much higher kinetic energy than the room temperature by several processes in plasmas, such as molecular dissociation and charge exchange with ions~\cite{Corrigan1965,Hey2004,McConkey2008}.
Such atoms collide each other distributing the injected energy to other atoms, and dissipate its energy to walls.

\Fref{sup:fig:md_temporal}~(b) and (c) show the results of the molecular-dynamics simulator \texttt{lammps}~\cite{LAMMPS} with $m=1$, $L=1$, $a=10^{-3}$, $r_0=10^{-3}$, $T_0=1$, $r=0.99$, $N=10^4$ and the time step of $2\times 10^{-5}$,
and with two different initial distributions, as similar to the Monte-Carlo simulation.
The values of $\langle E\rangle$ and $\langle E \rangle_\alpha$ are shown in \fref{sup:fig:md_temporal}~(b) and several snapshots of the energy distributions are shown in \fref{sup:fig:md_temporal}~(c).
While $\langle E\rangle$ decays to the steady-sate value, $\langle E \rangle_\alpha$ stays almost the same value.
The energy distribution has the power-law tail during the evolution, and the intensity and power-law index stays constant.
This suggests that the discussion in the main text does not rely on the details of the energy-dissipation process and applicable to wide variety of systems.

%% file: main.bbl
\providecommand{\noopsort}[1]{}\providecommand{\singleletter}[1]{#1}%
\begin{thebibliography}{32}%
\makeatletter
\providecommand \@ifxundefined [1]{%
 \@ifx{#1\undefined}
}%
\providecommand \@ifnum [1]{%
 \ifnum #1\expandafter \@firstoftwo
 \else \expandafter \@secondoftwo
 \fi
}%
\providecommand \@ifx [1]{%
 \ifx #1\expandafter \@firstoftwo
 \else \expandafter \@secondoftwo
 \fi
}%
\providecommand \natexlab [1]{#1}%
\providecommand \enquote  [1]{``#1''}%
\providecommand \bibnamefont  [1]{#1}%
\providecommand \bibfnamefont [1]{#1}%
\providecommand \citenamefont [1]{#1}%
\providecommand \href@noop [0]{\@secondoftwo}%
\providecommand \href [0]{\begingroup \@sanitize@url \@href}%
\providecommand \@href[1]{\@@startlink{#1}\@@href}%
\providecommand \@@href[1]{\endgroup#1\@@endlink}%
\providecommand \@sanitize@url [0]{\catcode `\\12\catcode `\$12\catcode
  `\&12\catcode `\#12\catcode `\^12\catcode `\_12\catcode `\%12\relax}%
\providecommand \@@startlink[1]{}%
\providecommand \@@endlink[0]{}%
\providecommand \url  [0]{\begingroup\@sanitize@url \@url }%
\providecommand \@url [1]{\endgroup\@href {#1}{\urlprefix }}%
\providecommand \urlprefix  [0]{URL }%
\providecommand \Eprint [0]{\href }%
\providecommand \doibase [0]{http://dx.doi.org/}%
\providecommand \selectlanguage [0]{\@gobble}%
\providecommand \bibinfo  [0]{\@secondoftwo}%
\providecommand \bibfield  [0]{\@secondoftwo}%
\providecommand \translation [1]{[#1]}%
\providecommand \BibitemOpen [0]{}%
\providecommand \bibitemStop [0]{}%
\providecommand \bibitemNoStop [0]{.\EOS\space}%
\providecommand \EOS [0]{\spacefactor3000\relax}%
\providecommand \BibitemShut  [1]{\csname bibitem#1\endcsname}%
\let\auto@bib@innerbib\@empty
\bibitem [{\citenamefont {Brilliantov}\ and\ \citenamefont
  {P{\"o}schel}(2004)}]{Brilliantov2004-dj}%
  \BibitemOpen
  \bibfield  {author} {\bibinfo {author} {\bibfnamefont {N.~V.}\ \bibnamefont
  {Brilliantov}}\ and\ \bibinfo {author} {\bibfnamefont {T.}~\bibnamefont
  {P{\"o}schel}},\ }\href@noop {} {\emph {\bibinfo {title} {Kinetic Theory of
  Granular Gases}}},\ Oxford Graduate Texts\ (\bibinfo  {publisher} {Oxford
  University Press},\ \bibinfo {address} {Oxford, New York},\ \bibinfo {year}
  {2004})\BibitemShut {NoStop}%
\bibitem [{\citenamefont {Aranson}\ and\ \citenamefont
  {Tsimring}(2006)}]{Aranson2006-qf}%
  \BibitemOpen
  \bibfield  {author} {\bibinfo {author} {\bibfnamefont {I.~S.}\ \bibnamefont
  {Aranson}}\ and\ \bibinfo {author} {\bibfnamefont {L.~S.}\ \bibnamefont
  {Tsimring}},\ }\href {\doibase 10.1103/RevModPhys.78.641} {\bibfield
  {journal} {\bibinfo  {journal} {Reviews of modern physics}\ }\textbf
  {\bibinfo {volume} {78}},\ \bibinfo {pages} {641} (\bibinfo {year}
  {2006})}\BibitemShut {NoStop}%
\bibitem [{\citenamefont {Ben-Naim}\ and\ \citenamefont
  {Machta}(2005)}]{Ben-Naim2005-rd}%
  \BibitemOpen
  \bibfield  {author} {\bibinfo {author} {\bibfnamefont {E.}~\bibnamefont
  {Ben-Naim}}\ and\ \bibinfo {author} {\bibfnamefont {J.}~\bibnamefont
  {Machta}},\ }\href {\doibase 10.1103/PhysRevLett.94.138001} {\bibfield
  {journal} {\bibinfo  {journal} {Physical review letters}\ }\textbf {\bibinfo
  {volume} {94}},\ \bibinfo {pages} {138001} (\bibinfo {year}
  {2005})}\BibitemShut {NoStop}%
\bibitem [{\citenamefont {Ben-Naim}\ \emph {et~al.}(2005)\citenamefont
  {Ben-Naim}, \citenamefont {Machta},\ and\ \citenamefont
  {Machta}}]{Ben-Naim2005-uz}%
  \BibitemOpen
  \bibfield  {author} {\bibinfo {author} {\bibfnamefont {E.}~\bibnamefont
  {Ben-Naim}}, \bibinfo {author} {\bibfnamefont {B.}~\bibnamefont {Machta}}, \
  and\ \bibinfo {author} {\bibfnamefont {J.}~\bibnamefont {Machta}},\ }\href
  {\doibase 10.1103/PhysRevE.72.021302} {\bibfield  {journal} {\bibinfo
  {journal} {Physical Review E}\ }\textbf {\bibinfo {volume} {72}},\ \bibinfo
  {pages} {021302} (\bibinfo {year} {2005})}\BibitemShut {NoStop}%
\bibitem [{\citenamefont {Kang}\ \emph {et~al.}(2010)\citenamefont {Kang},
  \citenamefont {Machta},\ and\ \citenamefont {Ben-Naim}}]{Kang2010-tk}%
  \BibitemOpen
  \bibfield  {author} {\bibinfo {author} {\bibfnamefont {W.}~\bibnamefont
  {Kang}}, \bibinfo {author} {\bibfnamefont {J.}~\bibnamefont {Machta}}, \ and\
  \bibinfo {author} {\bibfnamefont {E.}~\bibnamefont {Ben-Naim}},\ }\href
  {\doibase 10.1209/0295-5075/91/34002} {\bibfield  {journal} {\bibinfo
  {journal} {EPL}\ }\textbf {\bibinfo {volume} {91}},\ \bibinfo {pages} {34002}
  (\bibinfo {year} {2010})}\BibitemShut {NoStop}%
\bibitem [{\citenamefont {Bell}(1978)}]{Bell1978-wg}%
  \BibitemOpen
  \bibfield  {author} {\bibinfo {author} {\bibfnamefont {A.~R.}\ \bibnamefont
  {Bell}},\ }\href {\doibase 10.1093/mnras/182.2.147} {\bibfield  {journal}
  {\bibinfo  {journal} {Monthly notices of the Royal Astronomical Society}\
  }\textbf {\bibinfo {volume} {182}},\ \bibinfo {pages} {147} (\bibinfo {year}
  {1978})}\BibitemShut {NoStop}%
\bibitem [{\citenamefont {Gutenberg}\ and\ \citenamefont
  {Richter}(1944)}]{Gutenberg1944-wb}%
  \BibitemOpen
  \bibfield  {author} {\bibinfo {author} {\bibfnamefont {B.}~\bibnamefont
  {Gutenberg}}\ and\ \bibinfo {author} {\bibfnamefont {C.~F.}\ \bibnamefont
  {Richter}},\ }\href {\doibase 10.1785/bssa0340040185} {\bibfield  {journal}
  {\bibinfo  {journal} {Bulletin of the Seismological Society of America}\
  }\textbf {\bibinfo {volume} {34}},\ \bibinfo {pages} {185} (\bibinfo {year}
  {1944})}\BibitemShut {NoStop}%
\bibitem [{\citenamefont {Kolmogorov}(1991)}]{Kolmogorov1991-uo}%
  \BibitemOpen
  \bibfield  {author} {\bibinfo {author} {\bibfnamefont {A.~N.}\ \bibnamefont
  {Kolmogorov}},\ }\href@noop {} {\emph {\bibinfo {title} {Turbulence and
  Stochastic Process: Kolmogorov's Ideas 50 Years On}}},\ \bibinfo {type}
  {Tech. Rep.}\ (\bibinfo {year} {1991})\BibitemShut {NoStop}%
\bibitem [{\citenamefont {Hasegawa}\ and\ \citenamefont
  {Mima}(1977)}]{Hasegawa1977-oz}%
  \BibitemOpen
  \bibfield  {author} {\bibinfo {author} {\bibfnamefont {A.}~\bibnamefont
  {Hasegawa}}\ and\ \bibinfo {author} {\bibfnamefont {K.}~\bibnamefont
  {Mima}},\ }\href {\doibase 10.1103/PhysRevLett.39.205} {\bibfield  {journal}
  {\bibinfo  {journal} {Physical review letters}\ }\textbf {\bibinfo {volume}
  {39}},\ \bibinfo {pages} {205} (\bibinfo {year} {1977})}\BibitemShut
  {NoStop}%
\bibitem [{\citenamefont {Zipf}(1950)}]{Zipf1950-zy}%
  \BibitemOpen
  \bibfield  {author} {\bibinfo {author} {\bibfnamefont {G.~K.}\ \bibnamefont
  {Zipf}},\ }\href {\doibase
  10.1002/1097-4679(195007)6:3<306::aid-jclp2270060331>3.0.co;2-7} {\bibfield
  {journal} {\bibinfo  {journal} {Journal of clinical psychology}\ }\textbf
  {\bibinfo {volume} {6}},\ \bibinfo {pages} {306} (\bibinfo {year}
  {1950})}\BibitemShut {NoStop}%
\bibitem [{\citenamefont {Kraichnan}(1967)}]{Kraichnan1967-qf}%
  \BibitemOpen
  \bibfield  {author} {\bibinfo {author} {\bibfnamefont {R.~H.}\ \bibnamefont
  {Kraichnan}},\ }\href {\doibase 10.1063/1.1762301} {\bibfield  {journal}
  {\bibinfo  {journal} {The Physics of Fluids}\ }\textbf {\bibinfo {volume}
  {10}},\ \bibinfo {pages} {1417} (\bibinfo {year} {1967})}\BibitemShut
  {NoStop}%
\bibitem [{Note1()}]{Note1}%
  \BibitemOpen
  \bibinfo {note} {The direct molecular dynamics simulation for more realistic
  systems, the details of the probabilistic representation of Eq.~(\ref
  {eq:recursive}), and the detailed derivation several equations can be found
  in Supplemental Material, which includes Refs.~\cite
  {Ito1985-px,Corrigan1965,Hey2004,McConkey2008}}\BibitemShut {NoStop}%
\bibitem [{\citenamefont {Futcher}\ and\ \citenamefont
  {Hoare}(1980)}]{Futcher1980-ey}%
  \BibitemOpen
  \bibfield  {author} {\bibinfo {author} {\bibfnamefont {E.~J.}\ \bibnamefont
  {Futcher}}\ and\ \bibinfo {author} {\bibfnamefont {M.~R.}\ \bibnamefont
  {Hoare}},\ }\href {\doibase 10.1016/0375-9601(80)90042-0} {\bibfield
  {journal} {\bibinfo  {journal} {Physics letters. A}\ }\textbf {\bibinfo
  {volume} {75}},\ \bibinfo {pages} {443} (\bibinfo {year} {1980})}\BibitemShut
  {NoStop}%
\bibitem [{\citenamefont {Hendriks}\ and\ \citenamefont
  {Ernst}(1982)}]{Hendriks1982-cw}%
  \BibitemOpen
  \bibfield  {author} {\bibinfo {author} {\bibfnamefont {E.~M.}\ \bibnamefont
  {Hendriks}}\ and\ \bibinfo {author} {\bibfnamefont {M.~H.}\ \bibnamefont
  {Ernst}},\ }\href@noop {} {\bibfield  {journal} {\bibinfo  {journal} {Physica
  A: Statistical Mechanics and its Applications}\ }\textbf {\bibinfo {volume}
  {112}},\ \bibinfo {pages} {119} (\bibinfo {year} {1982})}\BibitemShut
  {NoStop}%
\bibitem [{\citenamefont {Futcher}\ and\ \citenamefont
  {Hoare}(1983)}]{Futcher1983-sf}%
  \BibitemOpen
  \bibfield  {author} {\bibinfo {author} {\bibfnamefont {E.~J.}\ \bibnamefont
  {Futcher}}\ and\ \bibinfo {author} {\bibfnamefont {M.~R.}\ \bibnamefont
  {Hoare}},\ }\href {\doibase 10.1016/0378-4371(83)90047-X} {\bibfield
  {journal} {\bibinfo  {journal} {Physica A: Statistical Mechanics and its
  Applications}\ }\textbf {\bibinfo {volume} {122}},\ \bibinfo {pages} {516}
  (\bibinfo {year} {1983})}\BibitemShut {NoStop}%
\bibitem [{\citenamefont {Herrmann}(2014)}]{Herrmann2014-ve}%
  \BibitemOpen
  \bibfield  {author} {\bibinfo {author} {\bibfnamefont {R.}~\bibnamefont
  {Herrmann}},\ }\href {\doibase 10.1142/8934} {\emph {\bibinfo {title}
  {Fractional calculus: An introduction for physicists}}}\ (\bibinfo
  {publisher} {World scientific},\ \bibinfo {year} {2014})\BibitemShut
  {NoStop}%
\bibitem [{\citenamefont {Anatolii Aleksandrovich~Kilbas}\ \emph
  {et~al.}(2006)\citenamefont {Anatolii Aleksandrovich~Kilbas}, \citenamefont
  {Srivastava},\ and\ \citenamefont
  {Trujillo}}]{Anatolii_Aleksandrovich_Kilbas2006-ts}%
  \BibitemOpen
  \bibfield  {author} {\bibinfo {author} {\bibfnamefont {A.}~\bibnamefont
  {Anatolii Aleksandrovich~Kilbas}}, \bibinfo {author} {\bibfnamefont {H.~M.}\
  \bibnamefont {Srivastava}}, \ and\ \bibinfo {author} {\bibfnamefont {J.~J.}\
  \bibnamefont {Trujillo}},\ }\href
  {https://play.google.com/store/books/details?id=LhkO83ZioQkC} {\emph
  {\bibinfo {title} {Theory And Applications of Fractional Differential
  Equations}}}\ (\bibinfo  {publisher} {Elsevier},\ \bibinfo {year}
  {2006})\BibitemShut {NoStop}%
\bibitem [{\citenamefont {Haubold}\ \emph {et~al.}(2011)\citenamefont
  {Haubold}, \citenamefont {Mathai},\ and\ \citenamefont
  {Saxena}}]{haubold_mittag-leffler_2011}%
  \BibitemOpen
  \bibfield  {author} {\bibinfo {author} {\bibfnamefont {H.~J.}\ \bibnamefont
  {Haubold}}, \bibinfo {author} {\bibfnamefont {A.~M.}\ \bibnamefont {Mathai}},
  \ and\ \bibinfo {author} {\bibfnamefont {R.~K.}\ \bibnamefont {Saxena}},\
  }\href {\doibase 10.1155/2011/298628} {\bibfield  {journal} {\bibinfo
  {journal} {Journal of Applied Mathematics}\ ,\ \bibinfo {pages} {Art. ID
  298628, 51}} (\bibinfo {year} {2011})}\BibitemShut {NoStop}%
\bibitem [{\citenamefont {Barabesi}\ \emph {et~al.}(2016)\citenamefont
  {Barabesi}, \citenamefont {Cerasa}, \citenamefont {Cerioli},\ and\
  \citenamefont {Perrotta}}]{barabesi_new_2016}%
  \BibitemOpen
  \bibfield  {author} {\bibinfo {author} {\bibfnamefont {L.}~\bibnamefont
  {Barabesi}}, \bibinfo {author} {\bibfnamefont {A.}~\bibnamefont {Cerasa}},
  \bibinfo {author} {\bibfnamefont {A.}~\bibnamefont {Cerioli}}, \ and\
  \bibinfo {author} {\bibfnamefont {D.}~\bibnamefont {Perrotta}},\ }\href
  {\doibase 10.1214/16-EJS1214} {\bibfield  {journal} {\bibinfo  {journal}
  {Electronic Journal of Statistics}\ }\textbf {\bibinfo {volume} {10}},\
  \bibinfo {pages} {3871} (\bibinfo {year} {2016})},\ \bibinfo {note}
  {publisher: Institute of Mathematical Statistics and Bernoulli
  Society}\BibitemShut {NoStop}%
\bibitem [{\citenamefont {Korolev}\ \emph {et~al.}(2020)\citenamefont
  {Korolev}, \citenamefont {Gorshenin},\ and\ \citenamefont
  {Zeifman}}]{korolev_mixture_2020}%
  \BibitemOpen
  \bibfield  {author} {\bibinfo {author} {\bibfnamefont {V.}~\bibnamefont
  {Korolev}}, \bibinfo {author} {\bibfnamefont {A.}~\bibnamefont {Gorshenin}},
  \ and\ \bibinfo {author} {\bibfnamefont {A.}~\bibnamefont {Zeifman}},\ }\href
  {\doibase 10.1007/s10958-020-04755-8} {\bibfield  {journal} {\bibinfo
  {journal} {Journal of Mathematical Sciences}\ }\textbf {\bibinfo {volume}
  {246}},\ \bibinfo {pages} {503} (\bibinfo {year} {2020})}\BibitemShut
  {NoStop}%
\bibitem [{\citenamefont {Fujii}(2022)}]{Fujii2022_PRE}%
  \BibitemOpen
  \bibfield  {author} {\bibinfo {author} {\bibfnamefont {K.}~\bibnamefont
  {Fujii}},\ }\href@noop {} {\bibfield  {journal} {\bibinfo  {journal}
  {Submitted to Physical Review E}\ } (\bibinfo {year} {2022})}\BibitemShut
  {NoStop}%
\bibitem [{\citenamefont {Massey}(1934)}]{Massey1934}%
  \BibitemOpen
  \bibfield  {author} {\bibinfo {author} {\bibfnamefont {H.~S.~W.}\
  \bibnamefont {Massey}},\ }\href {\doibase 10.1098/rspa.1934.0042} {\bibfield
  {journal} {\bibinfo  {journal} {Proceedings of the Royal Society of London.
  Series A, Containing Papers of a Mathematical and Physical Character}\
  }\textbf {\bibinfo {volume} {144}},\ \bibinfo {pages} {188} (\bibinfo {year}
  {1934})}\BibitemShut {NoStop}%
\bibitem [{\citenamefont {Flannery}(2006)}]{Flannery2006}%
  \BibitemOpen
  \bibfield  {author} {\bibinfo {author} {\bibfnamefont {M.}~\bibnamefont
  {Flannery}},\ }in\ \href {\doibase 10.1007/978-0-387-26308-3_45} {\emph
  {\bibinfo {booktitle} {Springer Handbook of Atomic, Molecular, and Optical
  Physics}}}\ (\bibinfo  {publisher} {Springer New York},\ \bibinfo {address}
  {New York, NY},\ \bibinfo {year} {2006})\ pp.\ \bibinfo {pages}
  {659--691}\BibitemShut {NoStop}%
\bibitem [{\citenamefont {Renyi}(2007)}]{Renyi2007-zw}%
  \BibitemOpen
  \bibfield  {author} {\bibinfo {author} {\bibfnamefont {A.}~\bibnamefont
  {Renyi}},\ }\href@noop {} {\emph {\bibinfo {title} {Probability Theory}}}\
  (\bibinfo  {publisher} {Courier Corporation},\ \bibinfo {year}
  {2007})\BibitemShut {NoStop}%
\bibitem [{\citenamefont {Tsallis}(1988)}]{Tsallis1988-ef}%
  \BibitemOpen
  \bibfield  {author} {\bibinfo {author} {\bibfnamefont {C.}~\bibnamefont
  {Tsallis}},\ }\href@noop {} {\bibfield  {journal} {\bibinfo  {journal} {J.
  Stat. Phys.}\ }\textbf {\bibinfo {volume} {52}},\ \bibinfo {pages} {479}
  (\bibinfo {year} {1988})}\BibitemShut {NoStop}%
\bibitem [{\citenamefont {Landsberg}\ and\ \citenamefont
  {Vedral}(1998)}]{Landsberg1998-lf}%
  \BibitemOpen
  \bibfield  {author} {\bibinfo {author} {\bibfnamefont {P.~T.}\ \bibnamefont
  {Landsberg}}\ and\ \bibinfo {author} {\bibfnamefont {V.}~\bibnamefont
  {Vedral}},\ }\href@noop {} {\bibfield  {journal} {\bibinfo  {journal} {Phys.
  Lett. A}\ }\textbf {\bibinfo {volume} {247}},\ \bibinfo {pages} {211}
  (\bibinfo {year} {1998})}\BibitemShut {NoStop}%
\bibitem [{\citenamefont {{Ito}}\ \emph {et~al.}(1985)\citenamefont {{Ito}},
  \citenamefont {{Tabata}}, \citenamefont {{Itoh}}, \citenamefont {{Morita}},
  \citenamefont {{Kato}},\ and\ \citenamefont {{Tawara}}}]{Ito1985-px}%
  \BibitemOpen
  \bibfield  {author} {\bibinfo {author} {\bibnamefont {{Ito}}}, \bibinfo
  {author} {\bibnamefont {{Tabata}}}, \bibinfo {author} {\bibnamefont
  {{Itoh}}}, \bibinfo {author} {\bibnamefont {{Morita}}}, \bibinfo {author}
  {\bibnamefont {{Kato}}}, \ and\ \bibinfo {author} {\bibnamefont {{Tawara}}},\
  }\href@noop {} {\bibfield  {journal} {\bibinfo  {journal} {IPPJ-AM41,
  Institute of Plasma}\ } (\bibinfo {year} {1985})}\BibitemShut {NoStop}%
\bibitem [{\citenamefont {Corrigan}(1965)}]{Corrigan1965}%
  \BibitemOpen
  \bibfield  {author} {\bibinfo {author} {\bibfnamefont {S.~J.~B.}\
  \bibnamefont {Corrigan}},\ }\href {\doibase 10.1063/1.1696701} {\bibfield
  {journal} {\bibinfo  {journal} {The Journal of Chemical Physics}\ }\textbf
  {\bibinfo {volume} {43}},\ \bibinfo {pages} {4381} (\bibinfo {year}
  {1965})}\BibitemShut {NoStop}%
\bibitem [{\citenamefont {Hey}\ \emph {et~al.}(2004)\citenamefont {Hey},
  \citenamefont {Chu}, \citenamefont {Mertens}, \citenamefont {Brezinsek},\
  and\ \citenamefont {Unterberg}}]{Hey2004}%
  \BibitemOpen
  \bibfield  {author} {\bibinfo {author} {\bibfnamefont {J.~D.}\ \bibnamefont
  {Hey}}, \bibinfo {author} {\bibfnamefont {C.~C.}\ \bibnamefont {Chu}},
  \bibinfo {author} {\bibfnamefont {P.}~\bibnamefont {Mertens}}, \bibinfo
  {author} {\bibfnamefont {S.}~\bibnamefont {Brezinsek}}, \ and\ \bibinfo
  {author} {\bibfnamefont {B.}~\bibnamefont {Unterberg}},\ }\href {\doibase
  10.1088/0953-4075/37/12/010} {\bibfield  {journal} {\bibinfo  {journal}
  {Journal of Physics B: Atomic, Molecular and Optical Physics}\ }\textbf
  {\bibinfo {volume} {37}},\ \bibinfo {pages} {2543} (\bibinfo {year}
  {2004})}\BibitemShut {NoStop}%
\bibitem [{\citenamefont {McConkey}\ \emph {et~al.}(2008)\citenamefont
  {McConkey}, \citenamefont {Malone}, \citenamefont {Johnson}, \citenamefont
  {Winstead}, \citenamefont {McKoy},\ and\ \citenamefont
  {Kanik}}]{McConkey2008}%
  \BibitemOpen
  \bibfield  {author} {\bibinfo {author} {\bibfnamefont {J.}~\bibnamefont
  {McConkey}}, \bibinfo {author} {\bibfnamefont {C.}~\bibnamefont {Malone}},
  \bibinfo {author} {\bibfnamefont {P.}~\bibnamefont {Johnson}}, \bibinfo
  {author} {\bibfnamefont {C.}~\bibnamefont {Winstead}}, \bibinfo {author}
  {\bibfnamefont {V.}~\bibnamefont {McKoy}}, \ and\ \bibinfo {author}
  {\bibfnamefont {I.}~\bibnamefont {Kanik}},\ }\href {\doibase
  10.1016/j.physrep.2008.05.001} {\bibfield  {journal} {\bibinfo  {journal}
  {Physics Reports}\ }\textbf {\bibinfo {volume} {466}},\ \bibinfo {pages} {1}
  (\bibinfo {year} {2008})}\BibitemShut {NoStop}%
\bibitem [{\citenamefont {Villani}(2006)}]{villani_mathematics_2006}%
  \BibitemOpen
  \bibfield  {author} {\bibinfo {author} {\bibfnamefont {C.}~\bibnamefont
  {Villani}},\ }\href {\doibase 10.1007/s10955-006-9038-6} {\bibfield
  {journal} {\bibinfo  {journal} {Journal of Statistical Physics}\ }\textbf
  {\bibinfo {volume} {124}},\ \bibinfo {pages} {781} (\bibinfo {year}
  {2006})}\BibitemShut {NoStop}%
\bibitem [{\citenamefont {Thompson}\ \emph {et~al.}(2022)\citenamefont
  {Thompson}, \citenamefont {Aktulga}, \citenamefont {Berger}, \citenamefont
  {Bolintineanu}, \citenamefont {Brown}, \citenamefont {Crozier}, \citenamefont
  {in~'t Veld}, \citenamefont {Kohlmeyer}, \citenamefont {Moore}, \citenamefont
  {Nguyen}, \citenamefont {Shan}, \citenamefont {Stevens}, \citenamefont
  {Tranchida}, \citenamefont {Trott},\ and\ \citenamefont {Plimpton}}]{LAMMPS}%
  \BibitemOpen
  \bibfield  {author} {\bibinfo {author} {\bibfnamefont {A.~P.}\ \bibnamefont
  {Thompson}}, \bibinfo {author} {\bibfnamefont {H.~M.}\ \bibnamefont
  {Aktulga}}, \bibinfo {author} {\bibfnamefont {R.}~\bibnamefont {Berger}},
  \bibinfo {author} {\bibfnamefont {D.~S.}\ \bibnamefont {Bolintineanu}},
  \bibinfo {author} {\bibfnamefont {W.~M.}\ \bibnamefont {Brown}}, \bibinfo
  {author} {\bibfnamefont {P.~S.}\ \bibnamefont {Crozier}}, \bibinfo {author}
  {\bibfnamefont {P.~J.}\ \bibnamefont {in~'t Veld}}, \bibinfo {author}
  {\bibfnamefont {A.}~\bibnamefont {Kohlmeyer}}, \bibinfo {author}
  {\bibfnamefont {S.~G.}\ \bibnamefont {Moore}}, \bibinfo {author}
  {\bibfnamefont {T.~D.}\ \bibnamefont {Nguyen}}, \bibinfo {author}
  {\bibfnamefont {R.}~\bibnamefont {Shan}}, \bibinfo {author} {\bibfnamefont
  {M.~J.}\ \bibnamefont {Stevens}}, \bibinfo {author} {\bibfnamefont
  {J.}~\bibnamefont {Tranchida}}, \bibinfo {author} {\bibfnamefont
  {C.}~\bibnamefont {Trott}}, \ and\ \bibinfo {author} {\bibfnamefont {S.~J.}\
  \bibnamefont {Plimpton}},\ }\href {\doibase 10.1016/j.cpc.2021.108171}
  {\bibfield  {journal} {\bibinfo  {journal} {Comp. Phys. Comm.}\ }\textbf
  {\bibinfo {volume} {271}},\ \bibinfo {pages} {108171} (\bibinfo {year}
  {2022})}\BibitemShut {NoStop}%
\end{thebibliography}%
